\documentclass[sigconf,screen,authorversion,nonacm]{acmart}
\usepackage{graphicx}
\usepackage{subcaption}
\AtBeginDocument{%
  }

\copyrightyear{2025}
\acmYear{2025}
\setcopyright{cc}
\setcctype{by-nc-nd}
\acmConference[ICCSIT 2025]{The 18th International Conference on Computer Science and Information Technology}{October 27--29, 2025}{Paris,, France}
\acmBooktitle{The 18th International Conference on Computer Science and Information Technology (ICCSIT 2025), October 27--29, 2025, Paris, France}
\acmPrice{}
\acmDOI{10.1145/3783862.3783879}
\acmISBN{979-8-4007-1858-8/2025/10}

\begin{document}

\title{A Brief History of the Waterfall Model: Past, Present, and Future}

\author{Antonios Saravanos}
\email{saravanos@nyu.edu}
\orcid{0000-0002-6745-810X}
\affiliation{%
  \institution{New York University}
  \streetaddress{7 East 12th Street, Room 625B}
  \city{New York}
  \state{NY}
  \country{USA}
  \postcode{10003}
}

\renewcommand{\shortauthors}{Saravanos, A.}

\begin{abstract}
The waterfall model, one of the earliest software development methodologies, has played a foundational role in shaping contemporary software engineering practices. This paper provides a historical and critical overview of the model, tracing its conceptual origins in software engineering, its formalization by Royce, and its evolution through decades of industry adoption and critique. Although often criticized for its rigidity, shortcomings, and high failure rates, the waterfall model persists in specific domains. Its principles continue to influence contemporary hybrid development frameworks that combine traditional and agile methods. Drawing on a range of scholarly sources, this study synthesizes key developments in the perception and application of the waterfall model. The analysis highlights how the model has shifted from a standalone framework to a component within modern hybrid methodologies. By revisiting its origins, assessing its present utility, and examining its role in contemporary development practices, this paper argues that the waterfall model remains relevant, not as a relic of the past but as part of context-aware development strategies. The paper contends that the model’s enduring relevance lies in its adaptability. By recognizing both its limitations and its strengths, and by understanding its integration within hybrid approaches, practitioners can make more informed decisions about methodology selection and process design in diverse development environments.
\end{abstract}

\ccsdesc[500]{Software and its engineering}
\ccsdesc[500]{Software and its engineering~Software creation and management}
\ccsdesc[500]{Software and its engineering~Software development process management}
\ccsdesc[500]{Software and its engineering~Software development methods}
\ccsdesc[500]{Software and its engineering~Waterfall model}

\keywords{Waterfall model, Systems development life cycle, Software development life cycle, SDLC, Diverse development environments}


\maketitle

\section{Introduction and Origins}

This paper offers a contemporary overview of the popular waterfall model. A structured formula for developing systems is often credited to Royce~\cite{ref37} (available in the ACM Digital Library), though some argue that it builds on earlier work by Benington~\cite{ref4}. Emerging during a period when the field was still in its formative stages, the model offered a much-needed structured approach to managing the growing complexity of software systems. It formalized a phase-driven development cycle that mirrored the logical progression of engineering projects, beginning with requirements gathering and ending in deployment and maintenance. As such, it became the blueprint for how software systems were conceptualized, designed, and built in the latter half of the twentieth century. The model gets its name from its visual appearance: a sequence of steps flowing downward from one phase to the next, resembling water flowing down a waterfall~\cite{ref40}. It is often categorized under the broader systems development life cycle (SDLC) umbrella, sometimes used interchangeably with the term software development life cycle~\cite{ref40}. Strictly speaking, one is part of the other~\cite{ref38}. As Ruparelia~\cite{ref38} explains, these distinctions have largely blurred in modern practice, where integrated systems development increasingly treats software as the central component. Consequently, the terms SDLC and waterfall model are frequently used synonymously in both academic and professional discourse~\cite{ref38}. According to Ruparelia~\cite{ref38}, ``a lifecycle covers all the stages of software from its inception with requirements definition through to fielding and maintenance''.

Royce’s 1970 formalization~\cite{ref37}, reprinted in 1987, comprises seven phases: system requirements, software requirements, analysis, program design, coding, testing, and operations, arranged so that each depends on the deliverables of the preceding one. His paper presents a sequence of refinements. The second model (a linear flow without feedback) and the third (with feedback to the prior phase) are the versions most commonly labeled as ``waterfall''. Notably, Royce recommended executing the development cycle at least twice, as illustrated in Figure~\ref{fig:waterfall-views}(b). The first pass, in his words, ``provides an early simulation of the final product''~\cite{ref37}, while the second produces a more robust solution. His paper was explicitly critical of a rigid, one-pass approach for large systems. Nevertheless, this guidance was often overlooked in early industry adoption, which favored the most linear interpretation and in turn invited later critiques of inflexibility and project failure. Contrary to popular belief, Royce did not advocate a strictly linear process; he prescribed repetition to accommodate learning and refinement. In retrospect, this emphasis on iteration and feedback foreshadows the incremental practices later emphasized by Agile methods.

Interestingly, the term ``waterfall'' does not appear in Royce’s~\cite{ref37} original paper. It was later popularized by Bell and Thayer~\cite{ref3} in 1976. Over the following decades, the waterfall model became widely institutionalized in both government and private-sector software development. It was often codified in official project management standards, particularly in regulated industries such as defense, aerospace, and healthcare. Despite its historical importance, the waterfall model has faced substantial criticism, especially since the 1990s. As the software industry encountered growing challenges with project overruns, shifting customer requirements, and rapid technological change, the model’s rigidity and limited adaptability came under scrutiny. Influential studies, including the Standish Group’s CHAOS Report~\cite{ref44}, reinforced these concerns by linking traditional approaches such as waterfall to high rates of project failure. These critiques, combined with the rise of iterative and incremental methods such as Agile, led many to question whether waterfall had outlived its usefulness.

Nonetheless, this paper argues that the waterfall model should not be dismissed as a relic of the past. Although it is no longer dominant as a standalone methodology, it continues to influence contemporary software engineering in important ways. First, it remains well suited to projects with stable requirements, clearly defined scopes, and strong demands for traceability and documentation. Second, its structure and discipline have gained renewed relevance in hybrid development methodologies that combine traditional and Agile practices. In these approaches, the sequential logic of waterfall is often retained at the macro level, such as in planning or compliance phases, while Agile methods are applied at the team or sprint level to enhance responsiveness and flexibility.

Building on this historical foundation, the remainder of this paper is organized as follows. Section~2 surveys the treatment of the waterfall model in mainstream software engineering literature, emphasizing how it has been interpreted, adapted, and critiqued. Section~3 examines its contemporary relevance, particularly its continued use both as a standalone method and as a component within hybrid methodologies. Section~4 presents a critical discussion and conclusion, reflecting on the enduring legacy and evolving applications of the waterfall model in modern software engineering.

\newlength{\waterfallwidth}
\newlength{\waterfallheight}
\setlength{\waterfallwidth}{0.3\textwidth}   
\setlength{\waterfallheight}{0.28\textwidth}  

\begin{figure*}[t]
    \centering

    \begin{subfigure}[t]{\waterfallwidth}
        \centering
        \includegraphics[width=\linewidth,height=\waterfallheight,keepaspectratio]{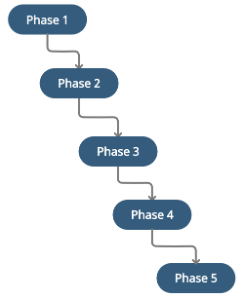}
        \caption*{(a)}
    \end{subfigure}%
    \hspace*{0.02\textwidth}%
    \begin{subfigure}[t]{\waterfallwidth}
        \centering
        \includegraphics[width=\linewidth,height=\waterfallheight,keepaspectratio]{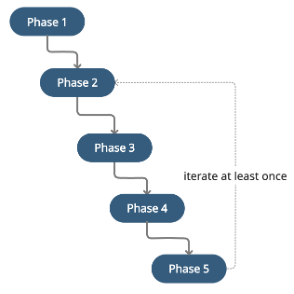}
        \caption*{(b)}
    \end{subfigure}

    \vspace{0.2em}

    \begin{subfigure}[t]{\waterfallwidth}
        \centering
        \includegraphics[width=\linewidth,height=\waterfallheight,keepaspectratio]{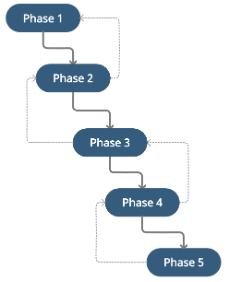}
        \caption*{(c)}
    \end{subfigure}%
    \hspace*{0.02\textwidth}%
    \begin{subfigure}[t]{\waterfallwidth}
        \centering
        \includegraphics[width=\linewidth,height=\waterfallheight,keepaspectratio]{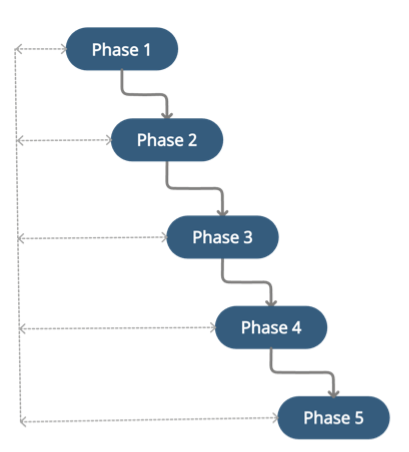}
        \caption*{(d)}
    \end{subfigure}

    \caption{Four Structural Views of the Waterfall Model: (a) Linear, (b) Two-Pass, (c) Feedback Loops to Previous Phase, (d) Feedback Loops to Any Previous Phase.}
    \label{fig:waterfall-views}
\end{figure*}

\section{Perceptions of Waterfall}

This section examines the evolving structure and perception of the waterfall model. First, we review how the model has been described and adapted in software engineering literature. We then examine its historical association with project failure.

\subsection{Composition of Waterfall}

As discussed in Section~1, the original waterfall model is often presented as a rigid, linear sequence of development phases. The literature, however, reveals a range of adaptations and interpretations. This subsection presents three representative snapshots from academic sources that illustrate how the model has been described over time. Notably, each of these sources depicts the waterfall model as consisting of five phases, rather than the seven originally described by Royce~\cite{ref37}.

The first example is Petersen et al.~\cite{ref30}, whose highly cited 2009 paper \emph{The Waterfall Model in Large-Scale Development} defines five phases. The first is requirements engineering, in which ``the needs of the customers are identified and documented on a high abstraction level'' and ``the requirements are refined so that they can be used as input to the design and implementation phase''~\cite{ref30}. The second is design and implementation, subdivided into two parts: the design, where ``the architecture of the system is created and documented'', and the implementation, where ``the actual development of the system takes place''~\cite{ref30}. The third phase is testing, in which ``the system integration is tested regarding quality and functional aspects''~\cite{ref30}. The fourth is release, defined as the point where ``the product is brought into a shippable state''~\cite{ref30}. The final phase is maintenance, where ``after the product has been released to the customer it has to be maintained'' and ``if customers discover problems in the product they report them to the company and get support in solving them''.

The second example is Sommerville’s~\cite{ref42} ninth edition of \emph{Software Engineering} (2011), which also presents five stages of the waterfall model. The first is requirements analysis and definition, in which ``the system’s services, constraints, and goals are established by consultation with system users'' and then ``defined in detail and serve as a system specification''~\cite{ref42}. The second is system and software design, where ``the systems design process allocates the requirements to either hardware or software systems by establishing an overall system architecture'', and ``software design involves identifying and describing the fundamental software system abstractions and their relationships''~\cite{ref42}. The third stage is implementation and unit testing, during which ``the software design is realized as a set of programs or program units'', each verified to ensure ``that each unit meets its specification''~\cite{ref42}. The fourth stage is integration and system testing, in which ``the individual program units or programs are integrated and tested as a complete system to ensure that the software requirements have been met'', after which ``the software system is delivered to the customer''~\cite{ref42}. Finally, operation and maintenance occurs when ``the system is installed and put into practical use'' and ``maintenance involves correcting errors which were not discovered in earlier stages of the life cycle, improving the implementation of system units, and enhancing the system’s services as new requirements are discovered''~\cite{ref42}. The same model appears in the tenth edition of the book, published in 2016~\cite{ref43}.

The third example is Andrei et al.~\cite{ref1}, who in 2019 describe the model with reference to Davis~\cite{ref9} (2012) and van Casteren~\cite{ref47} (2017). Their five stages are: requirements, defined as ``analyzing business needs and extensive documentation of all features''~\cite{ref1}; design, described as ``choosing all required technology and planning the full software infrastructure and interaction''~\cite{ref1}; coding, defined as ``solving all problems, optimizing solutions and implementing each component described in the requirements phase, using the diagrams and blueprints from the design phase''~\cite{ref1}; testing, described as ``extensive testing of all implemented features and components and solving any occurring issues''~\cite{ref1}; and finally operations, defined as ``deployment to a production environment''~\cite{ref1}. Table~\ref{tab:phases} summarizes how these three sources present the phases of the waterfall model.

\begin{table*}[t]
  \centering
  \caption{Phases of the waterfall model as presented by different sources.}
  \label{tab:phases}
  \begin{tabular}{@{}llll@{}}
    \toprule
    Phase & Petersen et al.\ \cite{ref30} (2009) & Sommerville \cite{ref42} (2011); \cite{ref43} (2016) & Andrei et al.\ \cite{ref1} (2019) \\
    \midrule
    1 & Requirements engineering & Requirements analysis and definition & Requirements \\
    2 & Design and implementation & System and software design & Design \\
    3 & Testing & Implementation and unit testing & Coding \\
    4 & Release & Integration and system testing & Testing \\
    5 & Maintenance & Operation and maintenance & Operations \\
    \bottomrule
  \end{tabular}
\end{table*}

We can see these phases connected in various ways across the literature, as illustrated in Figure~\ref{fig:waterfall-views}. The first waterfall lifecycle (a) is linear, with a single pass from start to finish. This representation is common in popular literature and is often associated with project failure, particularly in large projects. It reflects the second of the models presented by Royce~\cite{ref37}. The second model (b) illustrates a two-pass waterfall, where the requirements phase (Phase~1 in this example) is excluded from the second pass. This aligns with Andrei et al.~\cite{ref1}, who note that ``the Waterfall model assumes that once the initial requirements are set and every goal has been cleared of any ambiguities, there is an unobstructed road which the development team will follow towards finishing the project''. It should be emphasized that Royce~\cite{ref37} proposed several variations in conjunction with the two-pass idea, but since these are not commonly associated with the waterfall model, they fall outside the scope of this paper. The third model (c) represents a more contemporary interpretation of waterfall, where failure in one phase allows for feedback and repetition of the immediately preceding phase to correct mistakes. This corresponds to one of the variations described by Royce~\cite{ref37}. Finally, the fourth model (d) permits feedback to any earlier phase. This allows a team to revisit and repeat any previous stage in order to make corrections before moving forward.

\subsection{Reputation and Critique in Literature}

The waterfall model is closely associated with the notion of failure, a reputation it shares with the broader practice of software development~\cite{ref8,ref23}. Before the emergence of rapid application development (RAD) in 1991 and Agile in 2001, the software development life cycle and the waterfall model were largely synonymous. The idea of failure in software development was recognized early in the field’s history. Ebert~\cite{ref12} recalls that the 1968 conference, later recognized as the first on what became known as software engineering, identified ``the so-called software crisis'' as a central problem. Saravanos and Curinga~\cite{ref39} survey literature that highlights failure in the craft of software development. For example, Charette~\cite{ref34} observes that ``few IT projects, in other words, truly succeed''. Similarly, Bloch et al.~\cite{ref6} report that ``on average, large IT projects run 45 percent over budget and 7 percent over time, while delivering 56 percent less value than predicted'', based on a review of more than 5{,}400 IT projects each costing over \$15 million. Lauesen~\cite{ref27} notes that ``we have known for decades that IT projects often fail''. Charette~\cite{ref34} further states that ``from 5 to 15 percent'' of projects are ``abandoned before or shortly after delivery''. Gilb~\cite{ref14} critiques the model directly, calling it ``unrealistic, and dangerous to the primary objectives of any software project''. The most influential body of research linking software development with failure is the Standish Group’s CHAOS Report. Eveleens and Verhoef~\cite{ref19} cite the 1994 report, which stated that ``Standish reported a shocking 16\% project success rate, another 53\% of the projects had overruns of costs or time or less functionality and 31\% of the projects failed outright''. They emphasize the report’s impact on perception, writing that ``many authors have quoted the Standish figures to illustrate information technology is in a troublesome state'', that ``over the years their figures have had tremendous attention'', and that ``the figures indicate large problems with software engineering projects, and as such have had an enormous impact on information technology''. However, Cerpa and Verner~\cite{ref8} point out that Jørgensen and Moløkken-Østvold~\cite{ref20} questioned the methodology of the Standish Group, suggesting that ``there are serious problems with the way the Standish Group conducted their research and that the findings were biased toward reports of failure because a random sample of top IT executives was asked to share failure stories when mailed confidential surveys''. Consequently, it is possible that the reputation of failure was exaggerated, or at least not as severe as the reports suggested.

\section{Is There a Present and Future for the Waterfall Model?}

With the proliferation of Agile methodologies~\cite{ref13}, it is reasonable to ask whether the waterfall model still has a place in modern software engineering practice. Evidence shows that it continues to be used~\cite{ref18,ref30}. A 2019 study by Andrei et al.~\cite{ref1} reported that software developers applied waterfall 28.1\% of the time compared to Agile methods. Similarly, a 2020 survey conducted by the Project Management Institute (PMI)~\cite{ref31}, reported in its annual \emph{Pulse of the Profession} appendix~\cite{ref32}, found that slightly more than half of organizations (56\%) continued to use traditional approaches, including waterfall and similar structured methods (e.g., parallel or V-model).

Although Agile methods are gaining ground~\cite{ref28}, the persistence of waterfall suggests that it will remain relevant. Petersen et al.~\cite{ref30} note that ``the model is still widely used in software industry'', citing Raccoon~\cite{ref35} and adding that ``some researchers are even convinced that it will be around for a much longer period of time''. We agree and argue that waterfall remains relevant for two reasons: first, in projects where it aligns with project characteristics, and second, as a foundational component within emerging hybrid methodologies that integrate waterfall with other approaches (e.g., Agile, Scrum, iterative, and incremental).

Some projects continue to be well matched to the waterfall model. Mishra and Alzoubi~\cite{ref28} observe that ``many firms are still using the waterfall methodology since it simply works and has a successful track record''. In other words, project suitability is critical. Dennis et al.~\cite{ref10} also caution that ``choosing a methodology is not simple, because no single methodology is always best''. Wallis~\cite{ref48} identifies three strengths of waterfall. First is its clear structure, as ``the model provides a well-defined and structured approach to software development'', making it ``suitable for projects with stable and clearly defined requirements, where a sequential and linear development process is appropriate''~\cite{ref48}. Second is its focus on comprehensive documentation, an advantage in contexts requiring ``regulatory compliance, knowledge transfer, and future maintenance or enhancements''~\cite{ref48}. Third is its affinity for project planning, as it demands ``detailed project planning upfront'', which ``can be beneficial for managing resources, setting clear milestones, and estimating project timelines and costs''~\cite{ref48}. A common explanation for waterfall’s high failure rate is its misapplication to projects poorly suited for it. For example, some teams may use it simply because it is the only methodology they know. Wallis~\cite{ref48} warns that ``businesses should carefully consider whether the waterfall model aligns with their project requirements and organizational context'', noting that ``factors such as the stability of requirements, the need for flexibility, stakeholder involvement, and the dynamic nature of the industry should be evaluated''. For projects with these needs, Wallis~\cite{ref48} recommends Agile or iterative methodologies. Sommerville~\cite{ref42} similarly cautions that ``the waterfall model should only be used when the requirements are well understood and unlikely to change radically during system development''.

The second explanation for waterfall’s persistence is its incorporation into hybrid approaches. Kuhrmann et al.~\cite{ref24} define a hybrid approach as ``any combination of agile and traditional (plan-driven or rich) approaches that an organizational unit adopts and customizes to its own context needs''. Prenner et al.~\cite{ref33} similarly write that ``to benefit from the strengths of both approaches, software companies often use a combination of agile and plan-based methods, known as hybrid development approaches''. Tell et al.~\cite{ref45} provide a comparable definition: ``any combination of agile and traditional approaches that an organizational unit adopts and customizes to its own context needs''. The origins of hybrid practice can, according to Kirpitsas and Pachidis~\cite{ref21}, be traced to the work of Glass~\cite{ref16}, whose 2003 paper in \emph{IEEE Software} is often cited as an early reference. Küpper et al.~\cite{ref26}, drawing on the work of West et al.~\cite{ref50}, argue that ``hybrid software and systems development has become standard in practice''. The benefits of hybrid approaches are highlighted by Kuhrmann et al.~\cite{ref25}, who note that ``hybrid development provides a practical balance, combining the structure and predictability of traditional methods with the flexibility and responsiveness of agile approaches''. They further explain that ``these combinations are often not the result of deliberate planning but instead evolve organically based on practical experience, project needs, client demands, and regulatory requirements''~\cite{ref25}. This observation is echoed by Küpper et al.~\cite{ref26}, who reference the HELENA study~\cite{ref24} in stating that ``hybrid development approaches are barely planned or defined in advance''. Klünder et al.~\cite{ref22} similarly report that hybrid practices tend to emerge from a bottom-up rather than a top-down approach.

We can obtain an overview of the hybrid landscape as it relates to the waterfall model by examining a few key studies. Within the scope of this paper, we focus only on hybrid models that explicitly incorporate waterfall. The first is a systematic review conducted in 2020 by Prenner et al.~\cite{ref33}, who investigated how companies organize software development processes to combine Agile and plan-driven methods. Reviewing 24 papers, the authors concluded that all hybrid approaches fundamentally rely on the waterfall model, stating that ``all hybrid approaches are using in some way the phases described in Royce’s waterfall model''~\cite{ref33}. Prenner et al.~\cite{ref33} identified three organizational patterns: the waterfall--Agile approach (WAA), also called Agilefall, in which Agile methods are integrated into a waterfall structure; the waterfall--iterative approach (WIA), also called Waterative, where smaller waterfall cycles occur within iterations; and the pipeline approach (PA). Among these, WAA is the most widely used, followed by WIA and then PA. Combinations of approaches were also observed, such as WAA in conjunction with WIA.

A later study in 2022, a systematic literature review by Reiff and Schlegel~\cite{ref36}, provides ``a structured overview of the current state of research regarding the topic''. They identify two definitions of hybrid: first, ``a combination/mix of agile and traditional project management methodologies'', and second, ``the integration of an agile approach into existing traditional project management methodologies''~\cite{ref36}. The authors highlight four main hybrid models, two of which incorporate waterfall (water-scrum-fall and waterfall--Agile). They argue that hybrid approaches ``maximize project success'' and stress their value in allowing companies to ``use certain agile practices, even if there are constraints that impede the adoption of a pure agile approach''~\cite{ref36}. Reiff and Schlegel~\cite{ref36} conclude that ``hybrid systems that enable iteration and continuous evolution represent the future'' and call for further research to establish structured frameworks and more robust evaluations of hybrid project management methodologies. This reinforces the view that the waterfall model will persist, not as a standalone methodology, but as a component within hybrid approaches.

\section{Discussion and Conclusion}

As this paper has shown, the waterfall model holds a significant place in the evolution of software and systems development. From its conceptual roots in Benington’s~\cite{ref4} early process work to Royce’s~\cite{ref37} formalization, often misunderstood and oversimplified, the waterfall model shaped how developers and organizations approached complex projects. Although widely criticized for its rigidity and limited ability to accommodate changing requirements~\cite{ref49}, the sequential structure of the model continues to offer value in specific contexts, particularly where requirements are stable and well defined~\cite{ref10,ref11,ref42}. While the model has been closely linked to project problems and failures~\cite{ref15}, many of these outcomes can be attributed to misapplication. In particular, difficulties arise when waterfall is used in situations where requirements are unknown at the outset or subject to rapid change. This supports the pragmatic view expressed by Sommerville~\cite{ref41,ref42}, who advocates for context-sensitive methodology selection rather than adherence to a single universal model.

Despite the emergence~\cite{ref29}, rise~\cite{ref17,ref51}, and dominance of Agile (iterative and incremental) methodologies~\cite{ref7,ref13}, the waterfall model maintains a foothold in industry~\cite{ref1,ref13,ref30,ref31}. Recent shifts in software engineering show the model finding renewed purpose in hybrid approaches that blend waterfall with Agile, combining the strengths of both traditional practices (structure and rigor) and Agile practices (flexibility)~\cite{ref45}. Notable examples include Water-Scrum-Fall~\cite{ref46,ref50} and Scrumbanfall~\cite{ref5}, which demonstrate how waterfall principles have been selectively retained and integrated into modern development workflows. These developments suggest that the story of the waterfall model is not one of obsolescence but of evolution.

This historical and critical reflection underscores that the value of the waterfall model is not confined to the past. Its adaptability, whether through selective application or hybridization, points to an enduring relevance. The model continues to coexist alongside modern methodologies, with its core principles offering value in appropriate contexts. Future research should further examine contemporary uses of the waterfall model across projects of varying scales to extract lessons learned; refine simulation techniques to support evidence-based decisions around its use (see, for example, Bassil~\cite{ref2} and Saravanos and Curinga~\cite{ref39}); and contribute to the development of structured hybrid frameworks.

\end{document}